\documentclass[prl,aps,twocolumn]{revtex4}
\usepackage[]{times,amsmath,epsfig,graphicx}

\begin{document}

\title{On limitation of quality factor of single mode resonators with finite size}

\author{Fahmida Ferdous$^1$, Alena A. Demchenko$^2$, Sergey P. Vyatchanin$^2$,
Andrey B. Matsko$^{1}$, and Lute Maleki$^1$}

\affiliation{$^1$ OEwaves Inc., 465 North Halstead Street, Suite 140, Pasadena, CA 91107}

\affiliation{$^2$  Physics Department, Moscow State University, Moscow 119992 Russia}

\begin{abstract}
Using realistic numerical models we analyze radiative loss of bound and unbound modes of specially
designed high-Q whispering gallery and Fabry-Perot  cavities of similar size and shape, and find a set
of parameters when they can be treated as single mode structures. We show that these cavities have
similar properties in spite of their different loss mechanisms. The cavity morphology engineering
does not lead to reduction of the resonator quality factor.
\end{abstract}

\maketitle

An ideal optical cavity can be considered as a single dimension object with a well defined ``fundamental"
nearly-equidistant spectrum. The majority of realistic optical cavities
have large spectral density of modes of nearly identical quality (Q) factors where the fundamental
mode family is one of many \cite{boyd61bstj}. This is an undesirable feature for various practical applications. For
instance, high-order modes result in additional frequency dependent rejection or transmission of
a signal of interest if a cavity is used as a linear optical filter or a resonant sensor \cite{savchenkov05el,murugan11oe,ding12apl}. They also result in parametric instability in laser gravitational wave detectors \cite{01PLAbsv, 02PLAbsv,12UFNsvEn}.
High-order modes lead to mode competition in lasers \cite{sargentbook} and tend to cause unwanted nonlinear instabilities in optical sensors and oscillators. Different environmental sensitivity of the fundamental and high-order modes adds
complexity to the system because it becomes difficult to predict the spectrum of a cavity, especially a
monolithic one, with a reasonably large geometrical size, as a minute change of the ambient temperature
leads to a change in the cavity spectrum. In this paper we numerically study a method for reduction of
the complexity of the cavity frequency spectrum by optimizing the cavity morphology, using two distinct
examples of optical cavities -- whispering gallery mode (WGM) and Fabry-Perot (FP). We show that  very high-Q
single mode family cavities of both kinds are practically feasible.

The density of a cavity frequency spectrum can be reduced rather significantly by changing the cavity
morphology. One can get rid of the unwanted modes by properly shaping mirrors of an FP
\cite{pare92pra,kuznetsov05oe,tiffany07oe} or the circumference of a WGM
resonator \cite{savchenkov06ol,grudinin06oc}. The method draws from quantum mechanics permitting
existence of a potential well with a single bound state \cite{neamenbook}. Both FP and WGM resonators can be
approximated by a time-independent Schr\"odinger equation with a potential term determined by their
morphology. It is possible to create a single-mode family resonator since it is possible to select a
potential having only one bound state.

The method raises a fundamental question related to the lifetime limitation of the bound state. The
Schr\"odinger equation model does not tell much about the Q-factor of cavity modes. The lifetime of
the bound state is expected to be short due to various effects not taken into consideration. For example, diffraction loss becomes important
if the potential is too shallow and the system has finite overall dimensions comparable with the size of the
well. Increasing the well depth results in the increase of the lifetime, but it also leads to the eventual
confinement of higher order modes. In addition, the lifetime of light circulating in unbound modes
can still be significant, so the ``single-mode" feature becomes compromised. The question arises if it
is really possible to create a single mode cavity with high enough Q-factor. It is also
important to know what is the maximal ratio of the bound and unbound modes for a cavity of a finite
size. Finally, it is interesting if the Q-factor depends on the type of the cavity, since different
types of cavities have different loss mechanisms.
\begin{figure}[ht]
  \centering
  \includegraphics[width=8.5cm]{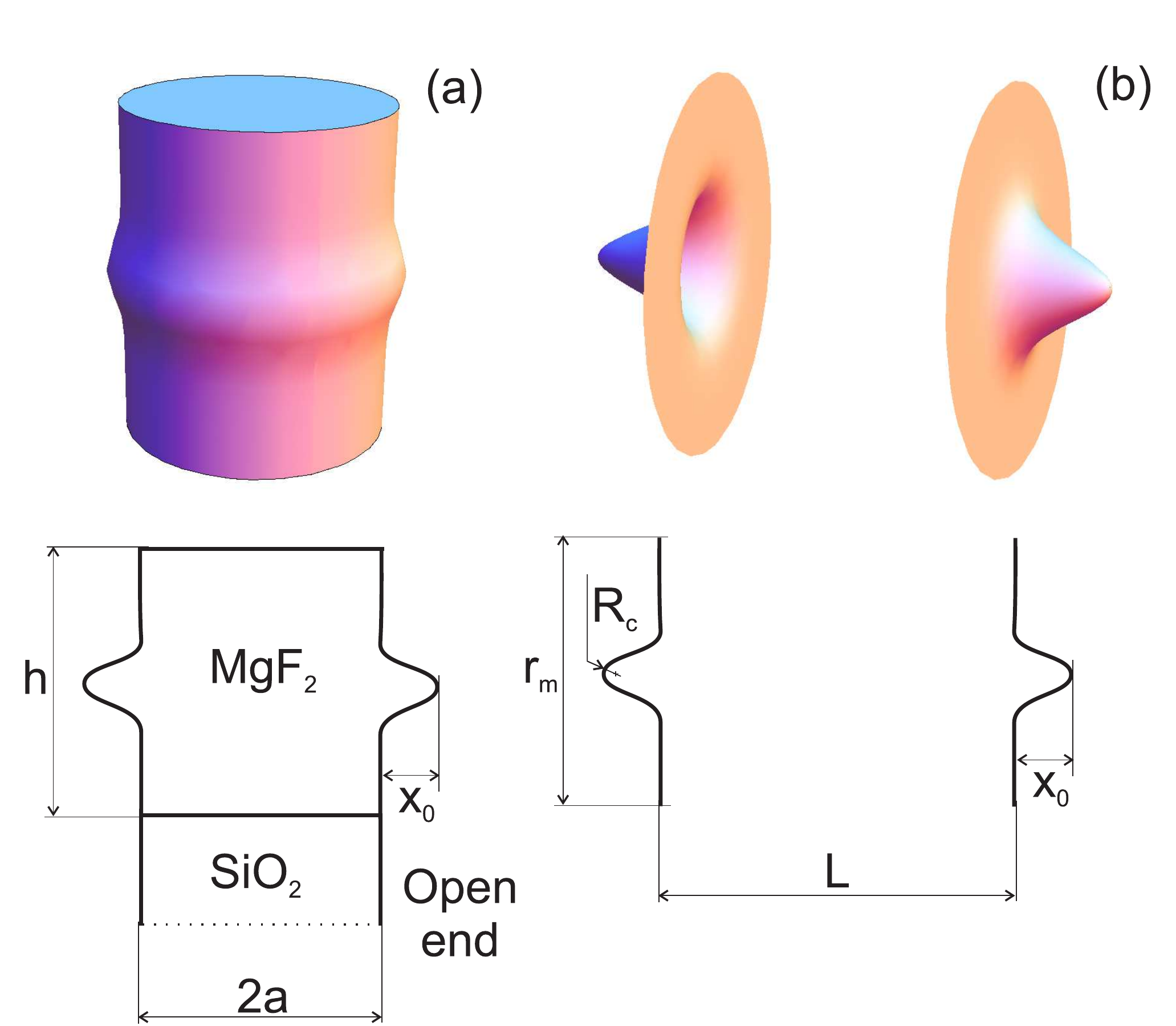}
\caption{ \small Morphologies of the studied cavities: (a) single-mode whispering gallery mode
resonator; (b) single-mode Fabry-Perot resonator. } \label{fig1}
\end{figure}

Loss in FP resonators having both bound and unbound modes has been previously evaluated in a theoretical study
\cite{tiffany07oe}. It was found that the bound modes have Q-factor that increase exponentially with the
mirror diameter, while the Q-factor of the unbound modes follows a power law dependence on the
diameter. Thus, this study shows that it is possible to achieve an infinite ratio of the Q-factors
for infinite mirrors. The study does not evaluate limitations imposed by the finite resonator size
when the mirror shape can be considered as a free parameter.

An analytical study of finite size single-mode resonators is rather challenging. In this paper we consider
realistic single mode FP and WGM cavities and study limitations of their Q-factors by means of
numerical simulations. For the sake of clarity we select resonators with 10~GHz free spectral range (FSR)
interrogated with $1.55~\mu$m wavelength light. The transverse dimension of resonators does not
exceed 0.1~cm. We confirmed that it is possible to make both single mode family FP and single mode
family WGM resonators by engineering their morphology, and found lifetime of light circulating within
the bound modes. The single-mode regime is achieved for a limited range of geometrical parameters of
the resonator. Q-factors of the single-mode resonator can exceed $Q=10^{10}$ and the maximum
ratio of the Q-factors of the bound and the highest Q unbound mode is approximately 1,000.

The Q-factor value of the FP resonator is comparable with the value of the diffraction-limited
Q-factor of a FP resonator with ideal spherical mirrors matched with the resonator length. For the
given resonator dimensions the Q-factor of the basic mode family is $Q=1.1\times 10^9$. The Q-factor is not
diffraction limited for a large (a few millimeters in diameter) overmoded WGM
resonator and the best practically achieved number exceeds $Q=10^{11}$. The single mode WGM resonator
has a comparable Q-factor in accordance with our simulations. These results imply that it is practical
to make these kinds of resonators for various applications.

We study a WGM resonator with morphology illustrated by Fig.~(\ref{fig1}a). A MgF$_2$ resonator with
thickness $h=60\ \mu$m and radius $a=0.35$~cm is considered. The resonator has 10~GHz FSR ($c/(2\pi a
n)$, $n_{WGM}=1.36$ is the refractive index of the material, $c$ is the speed of light in the
vacuum). The rim of the resonator has a Gaussian shape characterized with 10~$\mu$m full width at half
maximum (FWHM) and a variable height $x_0$.
\begin{equation}
x=x_0 \left \{ 1- \exp \left [-\left( \frac{t-h/2}{p} \right )^2 \right ] \right \},
\end{equation}
where $p^2=36.1\ \mu$m$^2$ corresponds to the FWHM selection. Using $x_0$ as a free parameter we find
lifetime of the cavity modes. The Gaussian profile is selected since this kind of profile was
demonstrated experimentally \cite{grudinin06oc}. The value of FWHM is not essential and the results
do not fundamentally change for different FWHM values. In this case, single mode operation is still observed at
different selection of $x_0$.

In accordance with the analytical model of a single-mode WGM resonator \cite{savchenkov06ol} there is
a restricted region of protrusion height, $x_0$, when only a single bound mode family exists in the
cavity. The theoretical model, though, requires the resonator thickness, $h$, to be infinite. The
model is not directly applicable to the case considered here since a change of the resonator shape
does not result in the change of the number of bound modes when the resonator has a fixed thickness.
All the modes have nearly identical high Q-factors in the resonator itself.

To address  this problem we place the WGM resonator on a SiO$_2$ substrate. The substrate
introduces additional attenuation to the resonator modes since its refractive index, $n_s=1.44$, is
larger as compared to the resonator host material. We equip the substrate with
antireflection coating to ensure that the light entering the substrate does not return to the WGMs.
\begin{figure}[ht]
  \centering
  \includegraphics[width=8.5cm]{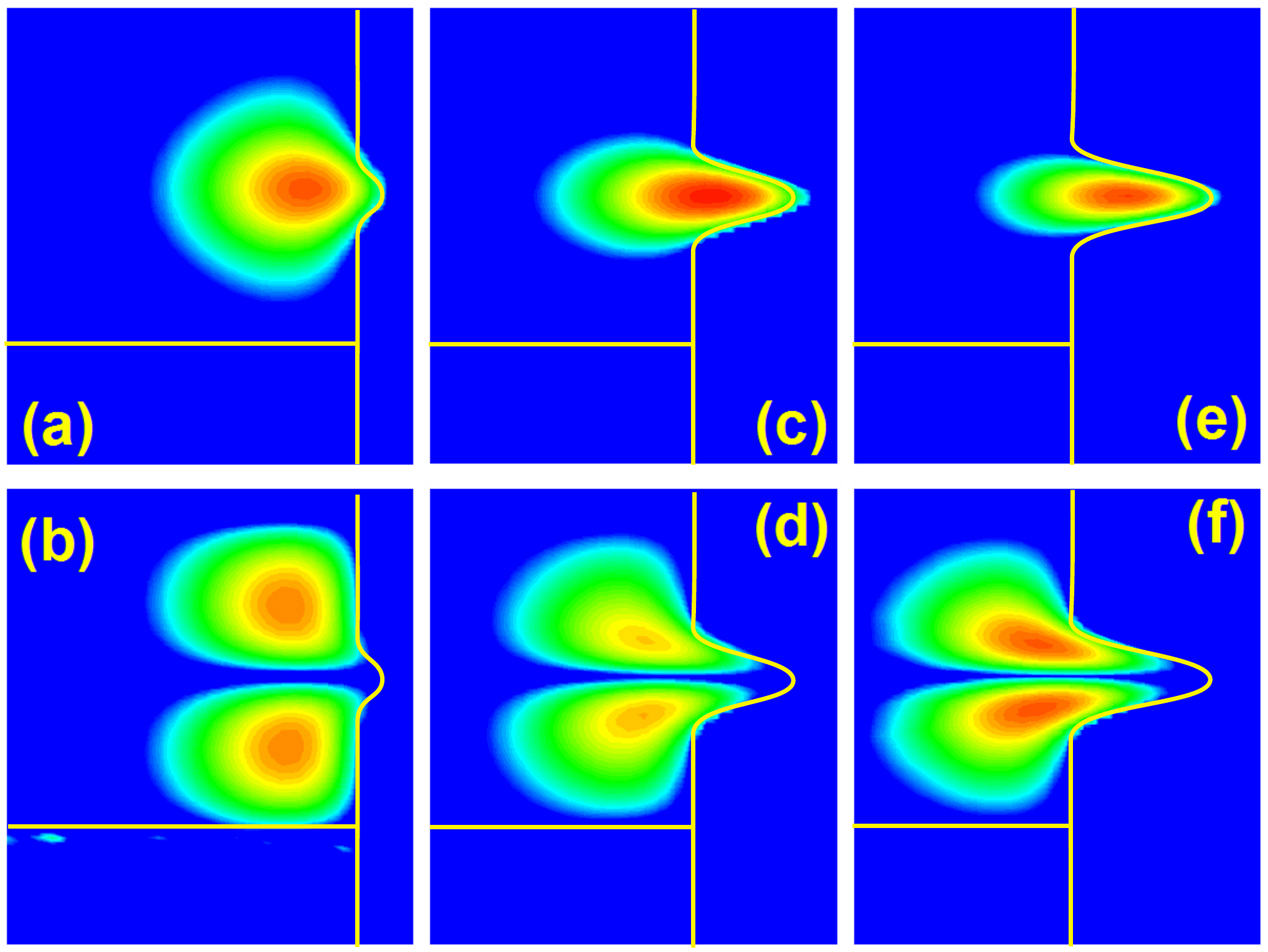}
\caption{ \small Spatial profile the field intensity for the fundamental and the first dipole WGMs
for different height of the protrusion: (a)~$x_0=2\ \mu$m, (b)~$x_0=8\ \mu$m, and (c)~$x_0=11\ \mu$m.
The intensity is illustrated by color density. } \label{fig2}
\end{figure}

Our approach utilizes COMSOL software to solve the Helmholtz equation using Galerkin's method of weighted residuals \cite{oxborrow07tmtt}. The 3D problem is reduced to a 2D problem because of azimuthal symmetry of the resonator modes.
Perfectly matched layers (PML) \cite{cheema13oe} are used to simulate power loss of the resonator modes originating from the mounting substrate. The layers block unwanted refections from the edges of the substrate taken into consideration in the code.  Because of the loss introduced by the PML, the frequencies of the modes ($\nu_0$) become complex and the Q-factor is calculated as $Q={\rm Re}\nu_0/(2 {\rm Im} \nu_0)$.

Results of the simulation show that light from both fundamental and first order dipole modes
escapes to the higher-index substrate, if the height of protrusion is small enough. The field leaving
the WGM is transferred to a Bessel beam freely propagating in the substrate
\cite{matsko05prl,savchenkov06oe}. Hence, the substrate can be considered as an ideal coupler for
generation of the Bessel beams out of the resonator.

The energy exchange efficiency reduces with the increase of the protrusion height. It also depends on the
mode order. There are two critical regimes corresponding to no bound states for $x_0=0$, and nearly
infinite number of high-Q bound states for $x_0 \gg \sqrt{2a \lambda}$ \cite{savchenkov06ol}. The consequence of the
protrusion increase is clearly seen at Fig.~(\ref{fig4}a) showing dependence of the Q-factor of the
fundamental and lowest order dipole WGMs. No modes are confined for small $x_0$. The first mode
becomes trapped as $x_0$ reaches a certain threshold value, equal to $2\ \mu$m for the selected
model. Its Q-factor grows exponentially, however it is still finite. The Q-factor of the first
dipole mode, on the other hand, barely changes. As the protrusion reaches another threshold point,
$x_0\simeq 8\ \mu$m, the Q-factor of the second mode starts to grow as well, which means that
the mode becomes bound.

The numerical simulation confirms the prediction of the simplified analytical theory showing that
there exist a certain renage of protrusion heights where the resonator can be treated as  single-mode
 \cite{savchenkov06ol}. It also shows that the maximum ratio of the Q-factors of the bound and
unbound modes is reached just before the second lowest order mode becomes bound. For the given
resonator geometry the ratio is approximately equal to a 1,000. The Q-factor of the fundamental
mode reaches its maximum at the same point. We find the maximum using interpolation since our method
of Q evaluation does not allow evaluating Qs exceeding several billions. The simulation does not
answer to the question related to the optimal resonator morphology to reach the highest possible
Q-factor under condition of the largest Q-factor difference. This problem needs further
investigation.

The reasoning related to the possibility of creation of a single mode family cavity seems to be
rather general and applicable to any kind of  cavity. To verify this intuitive conclusion we
consider a Fabry-Perot resonator shaped in a way shown in Fig.~(\ref{fig1}b) and having $L=1.5$~cm
length, $R_c=0.78$~cm radius of curvature, and $r_m=0.325$~mm radius of mirrors. The operational
wavelength is $\lambda=1.5\ \mu$m and the shape of the mirror is Gaussian, similarly  to the WGM
resonator
\begin{equation} \label{mirror}
h=x_0\left [ 1-\exp \left ( -\frac{r^2}{2 R_c x_0}  \right ) \right ].
\end{equation}
where $h$ is the deviation from the mirror plane, $r$ is the distance from the center of the mirror. The particular shape is selected so that the curvature radius of the mirror stays the same as the
optimal curvature of a spherical FP resonator with given dimensions. Our simulation shows that this
condition corresponds to the lowest loss of the fundamental mode.
\begin{figure}[ht]
  \centering
  \includegraphics[width=8.5cm]{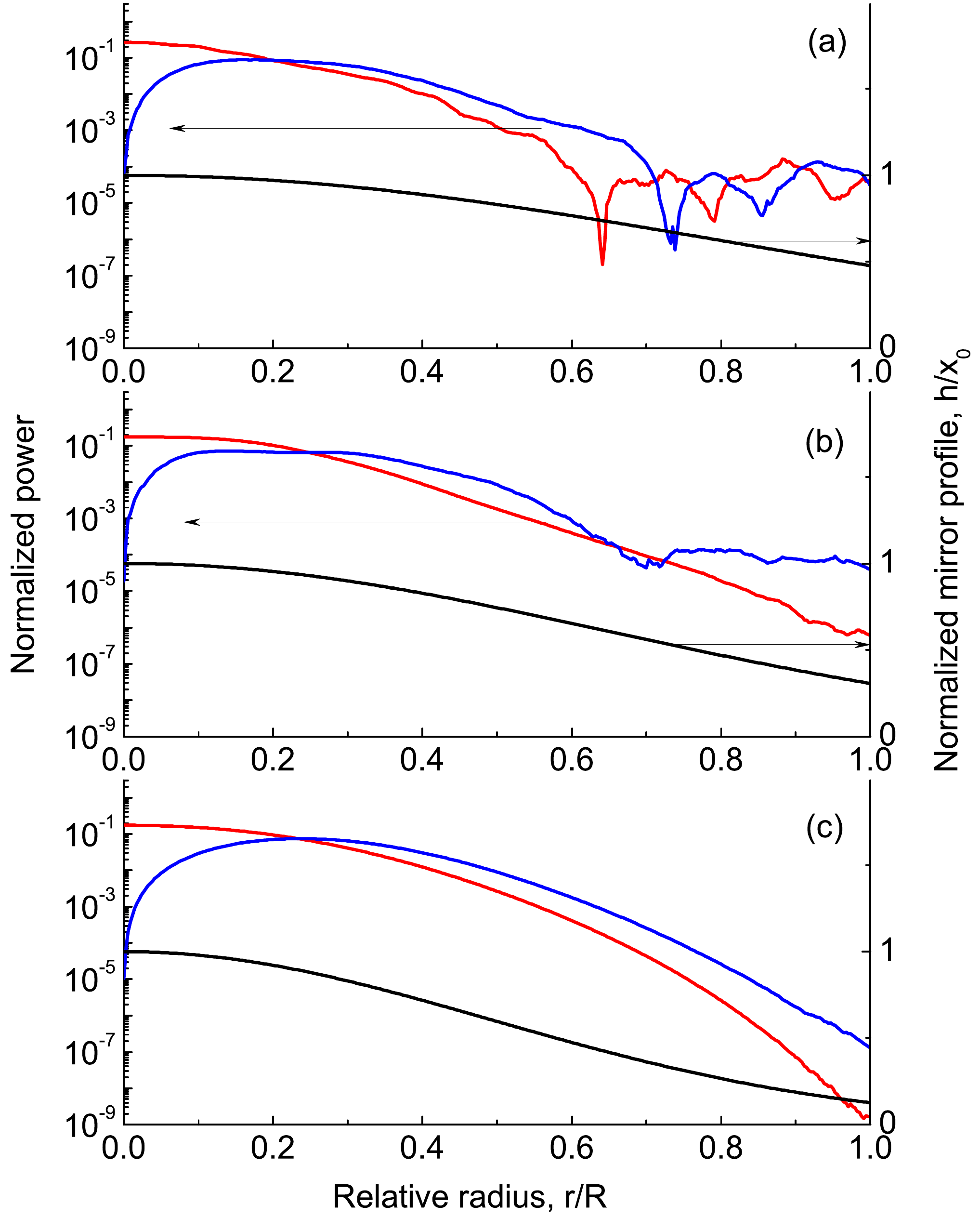}
\caption{ \small Spatial profile of the FP resonator mirrors and the field intensity for the
fundamental (red) and the first dipole (blue) modes for different heights of the mirrors: (a)$x_0=5\,\mu\text{m}$, (b) $x_0=8\,\mu\text{m}$, and
(c)$x_0=14\,\mu\text{m}$. The intensity distribution of poorly confined modes have irregular spatial structure
resulting from  diffraction at the mirror edge. } \label{fig3}
\end{figure}
\begin{figure}[ht]
  \centering
  \includegraphics[width=8.5cm]{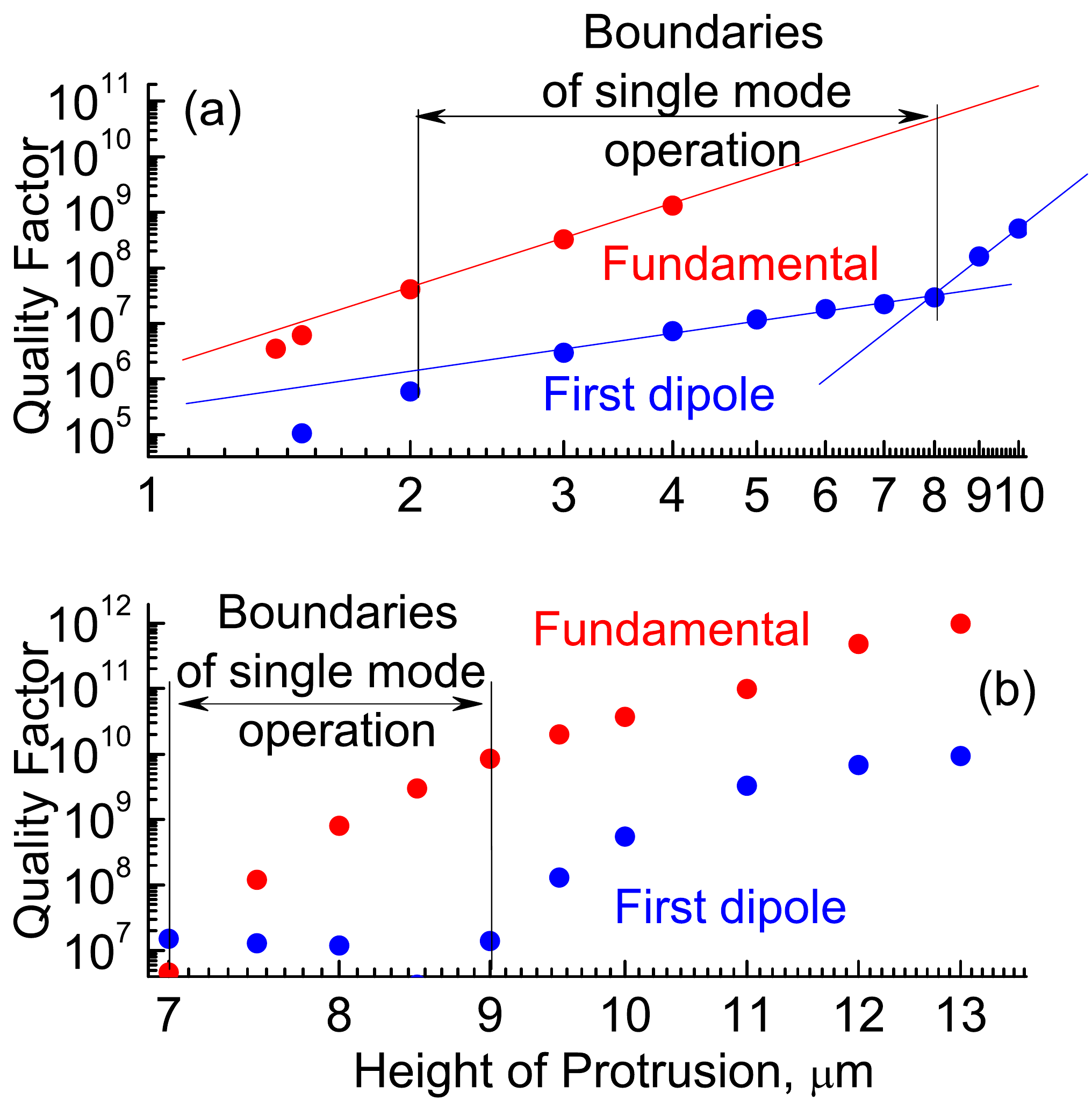}
\caption{ \small Dependence of the Q-factors of the fundamental and the first dipole modes on the
protrusion height for a whispering gallery mode resonator, (a), and Fabry-Perot resonator (b). The
numerical model is not accurate enough to predict Q-factors below $10^5$ and above $10^{10}$. }
\label{fig4}
\end{figure}

On the one hand, it is not obvious if the FP resonator has properties similar to those of the WGM
resonator, because the Q-factor of modes of the FP resonator itself is limited due to the
diffraction loss, while the WGM Q-factor is practically infinite for an isolated resonator. On the
other hand, both  WGM and FP resonators can be described in a similar manner if the Born-Oppenheimer
method is used, instead of paraxial approximation, to solve the Helmholtz equation (compare
\cite{savchenkov06ol} and \cite{nockel07oe}). The Born-Oppenheimer approach works in the case of the
nearly confocal FP cavity considered here ($2R_c \simeq L$). Since this approach results in 2D
Schr\"odinger equation and also allows studying different mirror shapes, we conclude that it is
possible to create a single mode FP resonator if we use the mirror profile given by Eq.~(\ref{mirror})
and adjust the height parameter $x_0$. Our numerical simulation (results shown in Figs.~\ref{fig3}
and \ref{fig4}b) confirms this conclusion. The FP resonator can be made with a single mode family.

Modes of a lossless FP cavity can be found from a set of scalar Fresnel integral equations \cite{siegman,solimeno}
\begin{eqnarray} \label{fp1}
\int G(\vec x_1,\vec x_2) \Psi_{2}(\vec x_2) d\vec x_2 = \xi \Psi_{1}(\vec x_1),\\ \label{fp2}
\int G(\vec x_1,\vec x_2) \Psi_{1}(\vec x_1) d\vec x_1 = \xi \Psi_{2}(\vec x_2),
\end{eqnarray}
where $\Psi_{1}$ and $\Psi_{2}$ stand for complex field amplitudes at the mirrors, $\xi=\exp (ikl)$, $k=2\pi/\lambda$ is the wave number, $l$ is an integer mode number, $d\vec x_{1,2} = x_{1,2} dx_{1,2} d\phi_{1,2}$, dimensionless coordinates relate to $r$ in \eqref{mirror} as $x_{1,2}=r_{1,2} \sqrt{k/L}$, $G(\vec x_1,\vec x_2)=-(i/2\pi) \exp \left [i((\vec x_1-\vec x_2)^2/2-y_1(\vec x_1)-y_2(\vec x_2) )  \right ] $ is the kernel, $y_{1,2}(\vec x_{1,2})= kh_{1,2}$ (see \eqref{mirror}) describe mirror profile.

We solved Eqs.(\ref{fp1},\ref{fp2}) numerically, assuming that the mirrors are identical ($y_1(\vec x)=y_2(\vec x)$) and demanding the field distribution on each mirror to approximately repeat itself after each round trip.  Since the complete repetition is impossible because of the diffraction loss we assumed that the spatial field profile of the field does not change, while its magnitude does, and introduced a complex scale factor $|\xi'| < 1$, instead of $\xi$, accounting for how much the field magnitude reduces from one round trip to another.  A discrete analog of the integral equations was solved using Hankel transform resulting in a considerable reduction of the evaluation time without accuracy loss (see \cite{vinet1993} for details). Among all solutions we selected those with $|\xi'|$ close to unity and found corresponding eigenfunctions. The relative power loss per round trip is given by $1-|\xi'|^2$, which allows finding finesse and Q-factor of the selected resonator modes.

The FP resonator has no bound modes if $x_0$ is small enough (Fig.~\ref{fig4}b). The fundamental mode
becomes bound when $x_0$ reaches $7\ \mu$m. The higher order modes are still unbound at this point.
An increase of $x_0$ results in the increase of Q-factor of the fundamental mode while the Q-factor of the other
modes does not change. The first dipole mode becomes bound for $x_0>9\ \mu$m, which is indicated by the
increase of its Q-factor. The maximum Q-factor of the fundamental mode reaches $10^{10}$ at this
point, and the ratio between Q-factors of the fundamental and the first dipole modes reaches a 1,000.

It is important to note here that the Q-factor of the single mode FP resonator exceeds that of the FP resonator with ideally matched spherical mirrors of identical dimensions. Our simulation shows that this effect occurs because the bound mode has super-Gaussian spatial profile. Our simulations show that the mode is stable with respect to the boundary conditions.

While this behavior is qualitatively similar to the behavior of the single mode WGM resonator
considered above, there are certain differences. The range of $x_0$ values required for the single
mode operation is different. The curvature of the FP mirror is important, while the curvature of the
WGM resonator protrusion is unimportant. The bound WGMs tend to have very high identical Q-factors,
while bound modes of a FP resonator have different Q-factors even for an ideally matched spherical
mirror. The protruded structure has much smaller dimensions as compared to those of the  single mode WGM resonator, while the spatial change of the mirror profile has comparable dimensions with the mirror size in the case of the FP resonator. The mirror deformation does not look as the exaggerated view of Fig.~(\ref{fig1}) (see mirror profile in Fig.~\ref{fig3}).

In conclusion, we have studied the dependence of the Q-factor of single mode whispering gallery mode and
Fabry-Perot cavities on their morphology and found it can be high enough to make the
resonator useful for various applications. These resonators perform in a similar way in spite of their
different structure.

Fahmida Ferdous, Andrey Matsko, and Lute Maleki acknowledge support from Defense Sciences Office of Defense Advanced Research Projects Agency under contract No. W911QX-12-C-0067 as well as support from Air Force Office of Scientific Research under contract No. FA9550-12-C-0068. Alena Demchenko and Sergey Vyatchanin acknowledge support from the Russian Foundation for Basic Research (Grant No. 14-02-00399A and Grant No. 13-02-92441 in frame of program ASPERA), and National Science Foundation (Grant No. PHY-130586).

\end{document}